\documentclass[reprint,prb,twocolumn,showpacs,superscriptaddress,aps,longbibliography]{revtex4-1}
\usepackage[utf8]{inputenc}

\usepackage{graphicx}
\DeclareGraphicsExtensions{.png,.jpg,.eps}

\usepackage{float}
\usepackage{xcolor}
\usepackage{amsmath}
\usepackage[printwatermark]{xwatermark}
\usepackage{float}
\usepackage[colorlinks=true,citecolor=blue]{hyperref}

\DeclareMathAlphabet\mathbfcal{OMS}{cmsy}{b}{n}

\newcommand{\lsoc} {\lambda_{\text{SOC}}}

\begin{document}

\title{Microscopic origin of multiferroic order in monolayer NiI$_2$}

\author{Adolfo O. Fumega}
\affiliation{Department of Applied Physics, Aalto University, 02150 Espoo, Finland}

\author{J. L. Lado}
\affiliation{Department of Applied Physics, Aalto University, 02150 Espoo, Finland}

\begin{abstract}

The discovery of multiferroic behavior in monolayer NiI$_2$ provides a new symmetry-broken state in van der Waals monolayers, featuring the simultaneous emergence of helimagnetic order and ferroelectric order at a critical temperature of $T=21$ K.  However, the microscopic origin of multiferroic order in NiI$_2$ monolayer has not been established, and in particular, the role of non-collinear magnetism and spin-orbit coupling in this compound remains an open problem.  Here we reveal the origin of the two-dimensional multiferroicity in NiI$_2$ using first-principles electronic structure methods.  We show that the helimagnetic state appears as a consequence of the long-range magnetic exchange interactions, featuring sizable magnetic moments in the iodine atoms. We demonstrate that the electronic density reconstruction accounting for the ferroelectric order emerges from the interplay of non-collinear magnetism and spin-orbit coupling. We demonstrate that the ferroelectric order is controlled by the iodine spin-orbit coupling, and leads to an associated electronically-driven distortion in the lattice.  Our results establish the microscopic origin of the multiferroic behavior in monolayer NiI$_2$, putting forward the coexistence of helical magnetic order and ligand spin-orbit coupling as driving forces for multiferroic behavior in two-dimensional materials.

\end{abstract}


\maketitle

\section{Introduction}

The emergence of symmetry-broken states in two-dimensional materials provides a unique opportunity for designing new forms of quantum matter in van der Waals heterostructures\cite{vdwHT2013}.  Well known examples are van der Waals monolayers hosting superconducting\cite{NbSe22015,Ising2018}, magnetic\cite{Fei2018,Huang2017,doi:10.1021/acs.nanolett.6b03052,Gong2017,doi:10.1021/acs.nanolett.9b00553}, and ferroelectric\cite{doi:10.1021/acs.nanolett.7b04852, Yuan2019} orders.  Interestingly, materials hosting two-different types of symmetry broken states allow for non-trivial couplings between different order parameters.  Multiferroics\cite{reviewMF2000,reviewMF2009,reviewMF2016} are a paradigmatic example of multiple symmetry breaking, where a material hosts simultaneous ferroelectric and magnetic orders. Two-dimensional multiferroics would provide disruptive possibilities for controlling magnetic order electrically\cite{switch2018,control2018}, and to design electrically switchable magnetic van der Waals heterostructures\cite{reviewmag2019,tunneling2018,VSe22020,Valley2018,Yurong2020,2021arXiv210311989V,Enhanced2017,TSC2020}.  In the pursuit of 2D multiferroic orders, a variety of families of two-dimensional materials provide potential candidate materials\cite{reviewmag2017,reviewmag2019,reviewmag2021}.  However, two-dimensional multiferroics have proven to be elusive.

Recent breakthrough experiments have shown the emergence of multiferroic symmetry-breaking in ultrathin NiI$_2$, leading to van der Waals materials displaying simultaneous magnetic and ferroelectric order. In particular, multiferroic behavior down to the bilayer,\cite{ctx1918398860006526} and ultimately to the monolayer limit\cite{song2021experimental}
was observed in NiI$_2$.  The monolayer of NiI$_2$ shows a magnetic transition temperature $T=21$ of K\cite{song2021experimental}, and at the very same temperature, a finite ferroelectric polarization emerges. In bulk compounds, a variety of mechanisms are known to drive multiferroic behavior\cite{PhysRevLett.96.067601,PhysRevLett.95.057205,PhysRevB.74.224444}, yet the origin of multiferroicity in monolayer NiI$_2$ remains an open question.

\begin{figure}
  \centering
  \includegraphics[width=\columnwidth]%
    {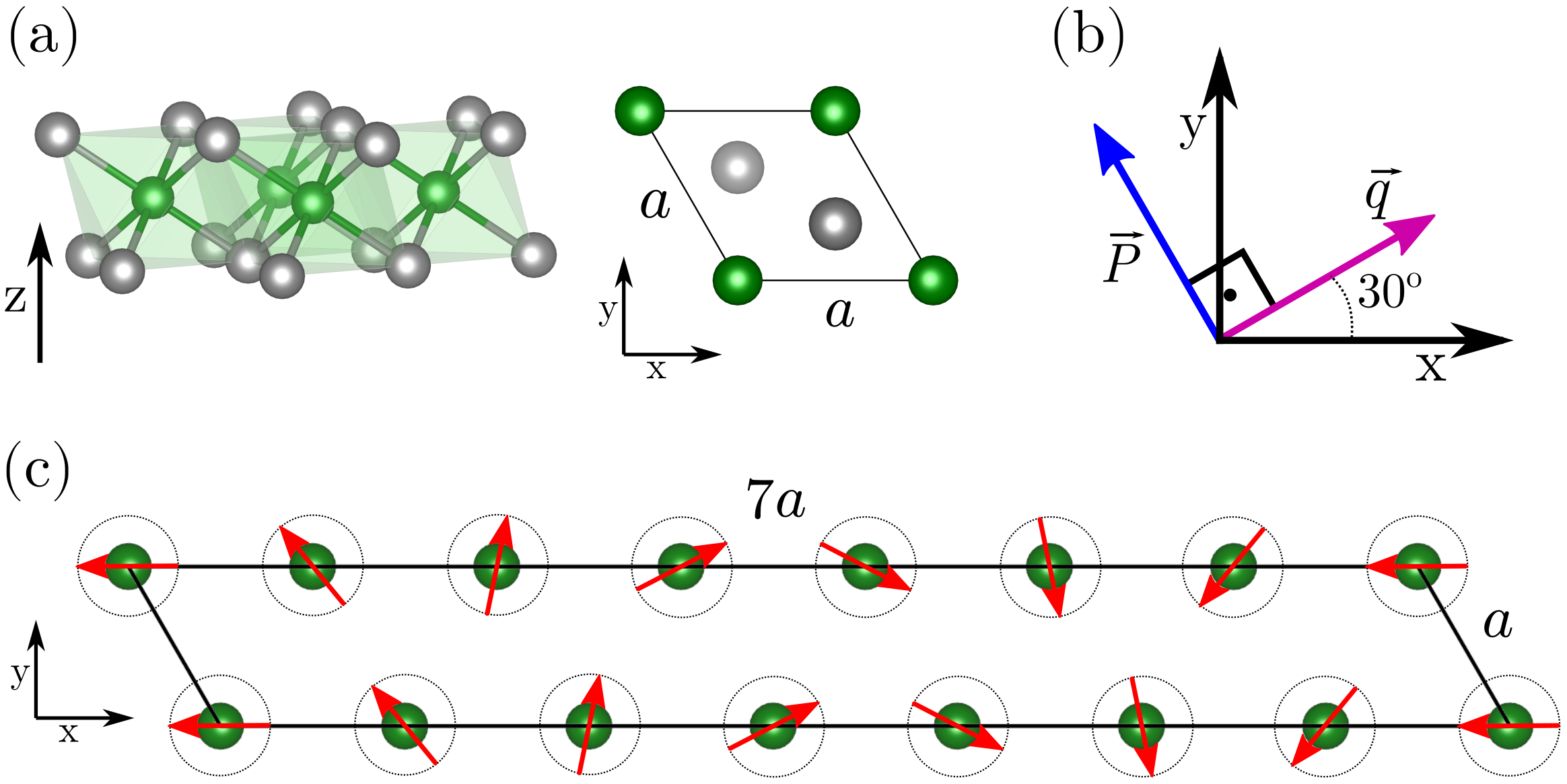}
     \caption{(a) Structure of monolayer NiI$_2$, showing top and side views. Ni atoms are depicted in green
     and I in gray. (b) Propagation direction in real space of the spin-spiral vector (pink) of the
     helimagnetic order and the induced ferroelectric polarization (blue). (c) Representation of the helimagnetic order, showing a periodicity of 
     $7\times 1$ unit cells.}\label{structs_mono}
\end{figure} 

Here we address the microscopic origin of multiferroicity in NiI$_2$ using first-principles methods.
In particular, we show that the combination of a helical magnetic state together with the
strong spin-orbit coupling of iodine leads to the emergence of a finite electric dipole.
Our results highlight the existence of a multiferroic state with a strong magnetoelectric coupling in the monolayer limit of NiI$_2$, and establish its origin in the combination of the non-collinear magnetic state and spin-orbit coupling.

\section{Computational methods}
We have performed \emph{ab initio} electronic structure calculations based on density functional theory\cite{HK} in NiI$_2$. Calculations were carried out with the all-electron full potential linearized augmented-plane-wave method, using a fully non-collinear formalism with spin-orbit coupling (SOC) as implemented in Elk\cite{Elk}. We have used the local density approximation (LDA) for the exchange-correlation functional\cite{KS}. The results presented are converged with respect to all the parameters, $2\times 14 \times 1$ k-mesh and a vacuum spacing of 20 \AA. 
Calculations of states with different helical states require careful convergence of the total electron density: with a $420 \times 60$ real space mesh for the electronic density and a convergence of the Kohn-Sham potential of $10^{-7}$ a.u..
Moreover, to see the effect of the SOC in the formation of the spontaneous electric polarization, we have performed calculations scaling both the overall strength of the SOC interaction
and the individual contribution of each atom by a dimensionless constant
$\lsoc$, where $\lsoc=1$ corresponds to the realistic limit.

\section{Helimagnetic state in the monolayer}

As a two-dimensional material, monolayer NiI$_2$ stems from a bulk van der Waals layered material.
In its bulk form, NiI$_2$ is formed by van der Waals layers in the 1T phase, like the one shown in Fig. \ref{structs_mono}a. Neutron diffraction has confirmed that this compound undergoes a first magnetic transition at $T_{N,1}\sim 76$ K to an antiferromagnetic state with ferromagnetic (FM) planes. A second transition to a helimagnetic (HM) state that displays a finite electric polarization is observed at low temperatures ($T_{N,2}\sim 59.5$ K).\cite{KUINDERSMA1981231, BILLEREY1977138}\footnote{The vector of the bulk
helimagnetic state is $\mathbf{q}=0.138 \mathbf{b}_1 + 1.457\mathbf{b}_3$,
where $\mathbf{b}_1$, $\mathbf{b}_3$
are the reciprocal lattice vectors}. 
Consequently, $T_{N,2}$ is not only the transition temperature to a HM, but to a multiferroic state
in the bulk crystal. In the monolayer limit, the transition to the magnetically ordered
states appears at $T_{N,2}=21$ K, appearing at the same temperature a ferroelectric polarization.
To unveil the origin of this multiferroic behavior,
we start analyzing the magnetic order of monolayer NiI$_2$. 

In order to model the helimagnetic order in the monolayer, we approximate the incommensurate spin vector by the closest
commensurate one $\textbf{q} = \frac{1}{7} \mathbf{b}_1$\footnote{
We neglect the out-of-plane component of the spin propagation
as we are in the monolayer limit and it is one order of magnitude smaller than the in-plane component.}
The commensurate in-plane spin vector $q-$vector differs only by a $\sim3.5$\% from the incommensurate one.
Therefore, the helimagnetic order is well captured in the monolayer with this spin arrangement.
In real space, our approximate spin propagation vector $\textbf{q}$ is displayed along the [210] direction (in lattice vector units), thus perpendicular to the [010] direction as shown in Fig. \ref{structs_mono}b. Then, the helimagnetic order in the monolayer can be modeled with a $7a\times a$ supercell and an in-plane spin cycloid as shown in Fig. \ref{structs_mono}c.

\begin{figure}[t!]
  \centering
  \includegraphics[width=\columnwidth]{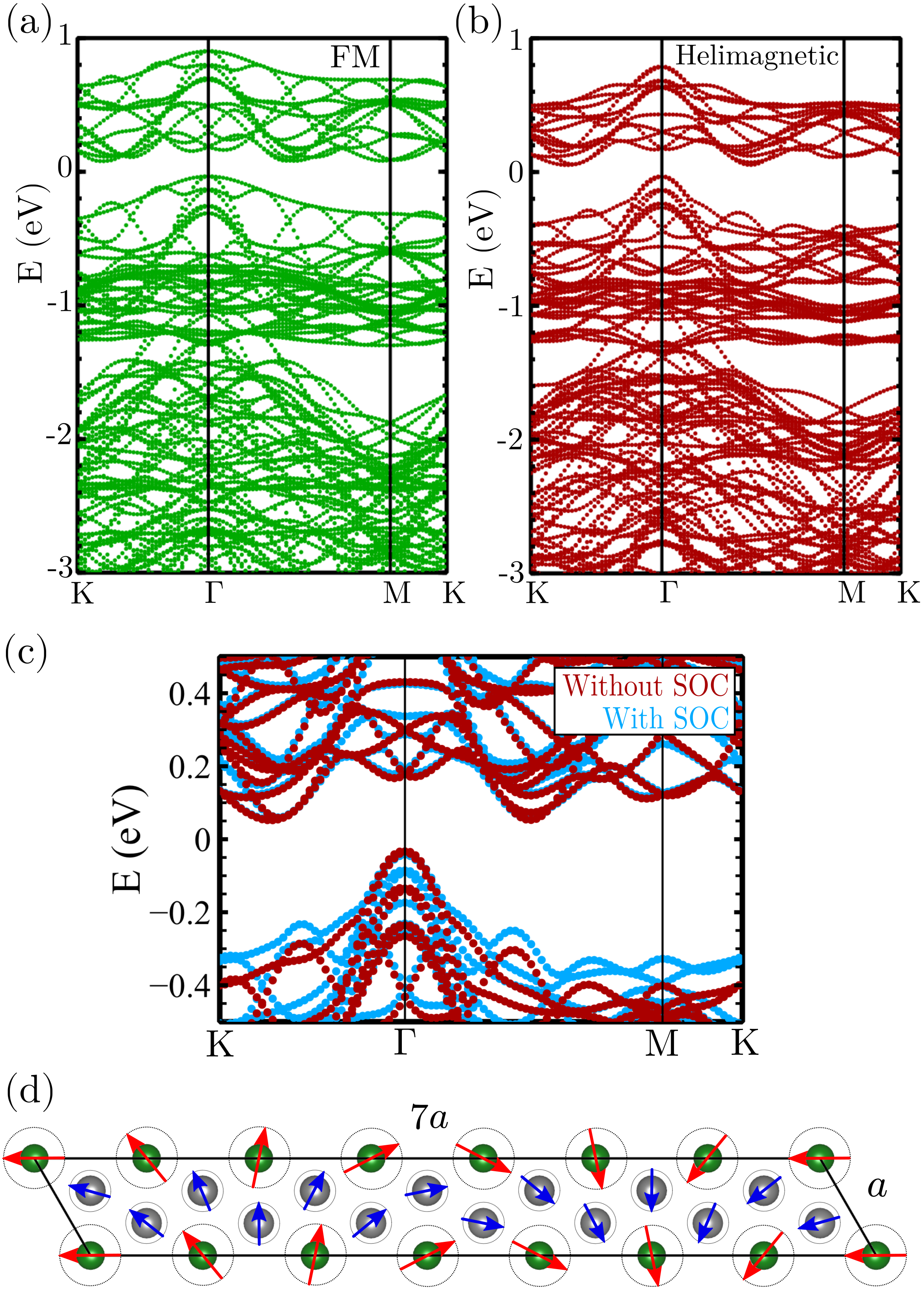}
     \caption{Band structure plots for the ferromagnetic
     (a) and helimagnetic (b) states.  A sizable energy gap can be seen
     for both configurations. (c) The effect of SOC in the helimagnetic state induces rearrangements in the band structure on the order of 50-100 meV. (d) Schematic of the magnetic polarization acquired by the I atoms in the helimagnetic state. Both the spin-vector direction and the helicity that occurs for the Ni atoms (red arrows) are transferred to the I atoms (blue arrows).}
     \label{bands}
     \label{Fig:bands}
\end{figure}

\begin{figure*}
  \centering
  \includegraphics[width=\textwidth]
        {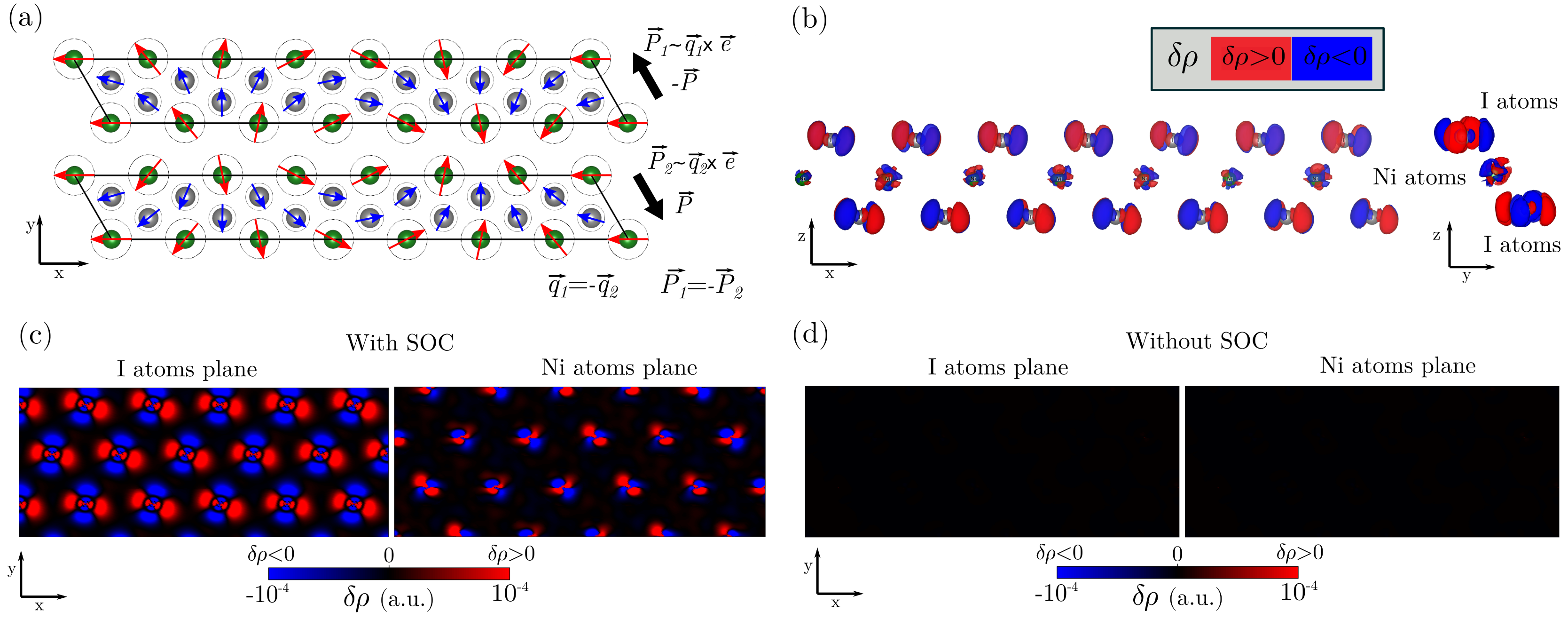}
     \caption{(a) Schematic of the two different spin spiral configurations with opposite $\mathbf{q}$ 
     used to test the emergence of the electric dipole in the presence of SOC. (b) Three dimensional plots from different spatial perspectives of the electronic density difference $\delta \rho$ for the helimagnetic configurations with opposite helicity when SOC is included.  
     This same quantity is shown as 2D plots in panels (c,d) for the planes of I and Ni atoms when the SOC is present (c) and absent (d). This unveils that an electronic reconstruction associated with the emergence of an spontaneous electric dipole is produced when SOC is introduced. It can be observed that the electronic rearrangement is larger around the I atoms than in the Ni ones. 
          In the absence of spin-orbit coupling, both electronic densities are identical
     as shown in panel (d), signaling a zero electric polarization (fully black).
     These results show that the electronic reconstruction leading
     to a ferroelectric dipole stem from the combination of the
     helimagnetic state spin-orbit coupling.}
     \label{elec_dens}
     \label{Fig:elec_dens}
\end{figure*}

We start addressing the electronic structure of the helimagnetic state.  From a chemical point of view, NiI$_2$ shows a 1T-phase with Ni atoms in a Ni$^{+2}$ state (d$^8$ S=1) in an octahedral iodine environment, leading to an effective $S=1$ triangular lattice model developing the spin spiral state. 
As a reference, it is interesting to compare the results of the helimagnetic state with a hypothetical ferromagnetic one.  The comparison between these magnetic configurations is performed without including SOC.  The band structure plots for both configurations are shown in Fig. \ref{bands}ab. A direct energy gap of $\sim 200$ meV coming from the $t_{2g}-e_{g}$ crystal field effect is found for both magnetic configurations.  The existence of a gap in both configurations demonstrates that the insulating origin stems from strong electronic correlations, and not the spin spiral itself \footnote{Due to a d-d character of the energy gap, electronic correlations are expected to increase the value predicted by LDA, thus approaching the experimental value reported for the bulk \cite{PhysRevB.35.4038}.}.  A comparison between the highest valence band of the helimagnetic and the FM configurations shows a decrease in the energies of the k-points for the HM. This is related to the energies found for each magnetic configuration.

The helimagnetic state is 27.7 meV per Ni atom lower in energy than the FM state. This result confirms that the helimagnetic state is, at least, a meta-stable low energy state at low temperature and, eventually, the ground state of 2D NiI$_2 $.  It is interesting to note that, from a fully ionic picture, the system would be ferromagnetic according to the Anderson-Goodenough-Kanamori FM superexchange\cite{PhysRev.79.350, GOODENOUGH1958287, KANAMORI195987} mediated by the $\sim 90^{\circ}$ Ni-I-Ni bonds, in stark contrast with experiments and our first-principles calculations\cite{KUINDERSMA1981231, BILLEREY1977138}. The covalent nature of this compound introduces long-range exchange interactions that lead to the helimagnetic state\cite{KUINDERSMA1981231}.  The compared energies of both configurations also show that the non-collinear magnetic order in the 2D limit arises from the long-range magnetic isotropic exchange interactions and the lattice frustration.
In particular, the relative sizes between first and long neighbor exchange determines the periodicity of the spin spiral order\cite{McGuire2017}. 
This is in agreement with previous analyses realized for the bulk \cite{KUINDERSMA1981231}. Therefore, our results highlight that the helimagnetic state does not rely on strong spin-orbit coupling effects\cite{DZYALOSHINSKY1958241,Amoroso2020,PhysRev.120.91} as its driving force.

We now move on to consider the effect of spin-orbit coupling in the electronic structure of the helimagnetic state. First, it is interesting to compare the electronic dispersion in the absence and presence of SOC, as shown in Fig. \ref{bands}c. In particular, it is clearly observed that states around the Fermi energy show variations on the order of 50-100 meV, well above the energy scale expected Ni spin-orbit coupling.  This phenomenology alone suggests that the spin-orbit coupling effects of iodine can be the dominating factor for states close to the Fermi energy.  It is worth noting that in order to obtain magnetic order at finite temperatures, spin-orbit coupling effects would provide the required SU(2) breaking terms in the Hamiltonian to overcome the Hohenberg-Mermin-Wagner theorem\cite{PhysRev.158.383,PhysRevLett.17.1133,CrI32017,halides2020}. The electronic reconstruction caused by the SOC will be related to the emergence of an electric dipole.
Furthermore, our first-principles results reveal that for the helimagnetic configuration, the iodine atoms develop a sizable magnetization of 0.23 $\mu_B$.  Therefore, due to the covalency of this compound, I atoms are substantially magnetized by the Ni atoms. Importantly, the helimagnetic state of the Ni atoms also occurs in the I atoms with the same spin propagation vector and helicity (see Fig. \ref{bands}d). This effect will be crucial for the discussion of the emergence of a spontaneous electric polarization in the helimagnetic state.

\section{Origin of ferroelectric order}

We now focus on analyzing the impact of spin-orbit coupling in the electronic structure,
and in particular in the electronic density of the system. As the electronic density of the
system determines the electric dipole, the mechanism for multiferroicity can be
directly inferred from change in the electronic density. 
From a Ginzburg–Landau perspective and symmetry considerations\cite{GL2015,PhysRevLett.96.067601},
the existence of a helical state allows for the emergence of a ferroelectric dipole
associated to the non-collinearity of the form

\begin{equation}\label{polarization}
    \mathbf P = \xi \mathbf q \times \mathbf e
\end{equation}
where $\mathbf P$ is the electric polarization, $\mathbf e = (0,0,1)$ the spin rotation axis and $\mathbf q_0 \approx \frac{1}{7} \mathbf b_1$ the $\mathbf{q}$-vector of the spin spiral, with $\mathbf b_1$ the reciprocal lattice vector of NiI$_2$, and $\xi$ is a scalar parameter that we will see that has a linear dependence with SOC and that has the same units as the electric polarization.  In particular, Eq. (\ref{polarization}) shows that two spin spiral configurations with $\mathbf q_1=\mathbf q_0$ and $\mathbf q_2 = -\mathbf q_0$ will give rise to opposite electric polarizations $\mathbf P_1 = - \mathbf P_2$ (Fig. \ref{Fig:elec_dens}a). The emergence of opposite electric dipoles can be directly observed in the total electronic density of the system. We perform first principles calculations in magnetic configuration with both $\mathbf{q}$-vectors, $\mathbf q_1$ and $\mathbf q_2$ (Fig. \ref{Fig:elec_dens}a).  Both configurations are energetically equivalent, and therefore show same energies with and without spin-orbit coupling.

\begin{figure}[t!]
  \centering
  \includegraphics[width=0.85\columnwidth]%
    {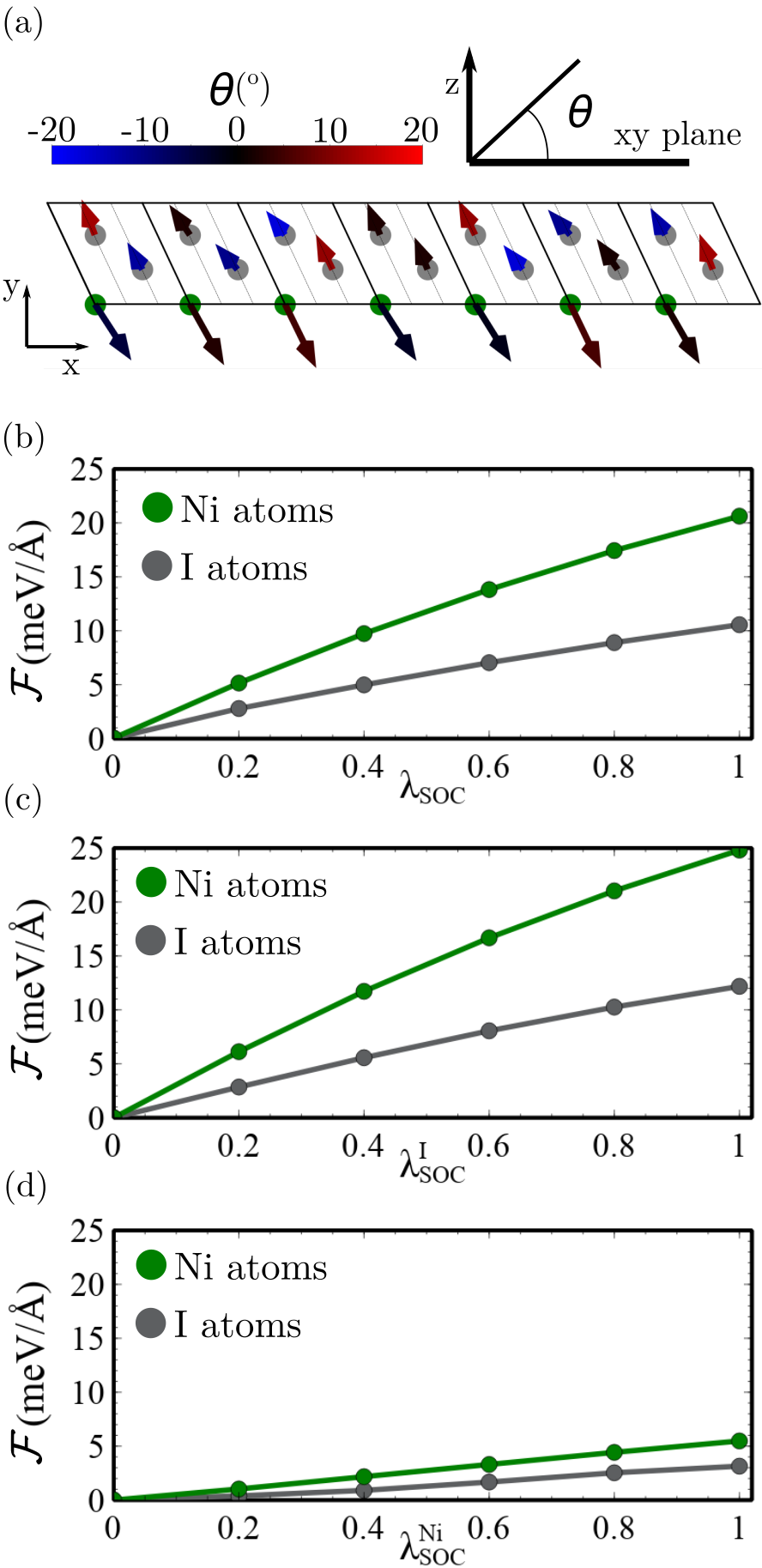}
     \caption{(a) Real space plot
     of the ferroelectric force
     $\mathbfcal{F}_\alpha$, highlighting
     the atomic relaxations associated
     to the emergence of a finite electric
     polarization.
     Panel (b) shows the
     scaling of the total
     ferroelectric force
     as a function of the SOC dimensionless parameter $\lambda_{SOC}$,
     highlighting the SOC origin
     of ferroelectric relaxations.
     Panel (c,d) show the ferroelectric force
     as a function of (c) the iodine SOC $\lsoc^{\text{I}}$,
     taking $\lsoc^{\text{Ni}}=0$, and (d) 
     as a function of the nickel SOC $\lsoc^{\text{Ni}}$,
     taking $\lsoc^{\text{I}}=0$. This highlights the
     dominant role of iodine in generating the
     ferroelectric order and the opposite sign of the nickel contribution.}
     \label{forces}
     \label{Fig:forces}
\end{figure}

To reveal the emergence of the electric dipole, we now analyze the difference in the total electronic densities $\rho_1$ and $\rho_2$, associated with ground states with $\mathbf q_1$ and $\mathbf q_2$, respectively.  In particular, the difference between these two electronic densities directly shows in real space the origin of the associated ferroelectric dipole, as the configurations will show opposite ferroelectric polarizations $\mathbf P_1 = -\mathbf P_2$.  First, we show in Fig. \ref{Fig:elec_dens}b the electronic difference $\delta \rho = \rho_1 - \rho_2$ in real space between the two spin-spiral configurations with spin-orbit coupling. In particular, it is clearly observed that the changes in the density are maximal in the iodine atoms, highlighting the dominant role of iodine in the rearrangement of the electronic charge, ultimately leading to an electric dipole.  The dominant role of iodine is also observed by showing the contribution of the iodine and Ni to the electronic reconstruction as shown in Fig. \ref{Fig:elec_dens}c, showing that the reconstruction is maximal in the iodine plane.  Finally, we address the case without spin-orbit coupling, shown in Fig. \ref{Fig:elec_dens}d. In this case, it is clearly observed that the charge density is the same for both spin spirals, leading to a zero contribution to the electric dipole. The previous results highlight that the combination of helical order and spin-orbit coupling is responsible for the charge reconstruction leading to an electric dipole.

The previous discussion was focused on the electronic contribution to the electric polarization.  Nevertheless, apart from this electronic contribution, the rearrangement of the electronic density will also have an impact on the lattice, leading to a small atomic relaxation associated with the ferroelectric electronic density. This lattice contribution can be directly imaged by comparing the small lattice relaxations associated with the helical orders $\mathbf{q}_1$ and $\mathbf{q}_2$. In particular, we now characterized the ferroelectric atomic relaxation by the component in the forces associated with the two configurations with opposite ferroelectric vectors as

\begin{equation}
    \mathbfcal{F}_\alpha = 
    \frac{1}{2} ( 
    \mathbf F_\alpha [\mathbf q_1] -
    \mathbf F_\alpha [\mathbf q_2]
    )
\end{equation}
where $\mathbf F_\alpha [\mathbf q_1]$, $\mathbf F_\alpha [\mathbf q_2]$ is the total force for atom $\alpha$ for the spin spiral state $\mathbf q_1$, $\mathbf q_2$ respectively \footnote{We take as initial structure the fully relaxed structure in the minimal unit cell.}.  In the absence of spin-orbit coupling, the ferroelectric force $\mathbfcal{F}_\alpha$ is exactly zero, consistent with the vanishing difference in the electronic densities shown in Fig. \ref{Fig:elec_dens}d.  In stark contrast, in the presence of the helical order and spin-orbit coupling, the emergent electronic reconstruction leads to opposite atomic relaxations for $\mathbf q_1$ and $\mathbf q_2$, yielding a finite $\mathbfcal{F}_\alpha$.  We show in Fig. \ref{Fig:forces}a the ferroelectric force acting in the different iodine and Ni atoms. It is clearly observed that the ferroelectric polarization leads to an atomic displacement in the lattice. The previous results highlight that the combination of spin helical order and spin-orbit coupling leads not only to a ferroelectric order, but also to a strong magneto-elastic coupling.

We finally analyze the spin-orbit origin of the
ferroelastic coupling. For this purpose, we now divide the density
functional Hamiltonian into the relativistic and non-relativistic
part as

\begin{equation}
\mathcal{H} = 
\mathcal{H}_0 +
\lsoc \mathcal{H}_{\text{SOC}}
\end{equation}

where $\mathcal{H}_0$ is the non-relativistic Hamiltonian and $\mathcal{H}_{SOC}$ is the relativistic spin-orbit coupling contribution.  The parameter $\lsoc$ provides a knob to controllably switch on spin-orbit coupling, where $\lsoc=0$ corresponds to the non-relativistic calculations, and $\lsoc=1$ corresponds to the real spin-orbit coupling.  We show in Fig. \ref{Fig:forces}b the evolution of the average module of the ferroelectric force as a function of the spin-orbit parameter, computed for the two atomic species $\mathcal{F}_{\text{Ni}} = \langle | \mathbfcal{F}_\alpha | \rangle_{\text{Ni}}$ and $\mathcal{F}_{\text{I}} = \langle | \mathbfcal{F}_\alpha | \rangle_{\text{I}}$.  In particular, it is clearly observed that the spiral-induced distortion of the lattice is zero in the absence of spin-orbit coupling, and it increases approximately linearly with the spin-orbit coupling, thus confirming the linear dependence with SOC anticipated for the $\xi \sim \lsoc$ parameter in Eq. (\ref{polarization}). 
This phenomenology shows that spin-orbit coupling is central in the induced ferroelastic coupling, in agreement with the electronic density reconstruction observed.

The previous ferroelectric force triggers a distortion of the lattice sites $d$ on the order of $0.5$pm, yielding a lattice contribution to the ferroelectric order. 
We can provide an estimate of the electric dipole in the monolayer $p_{mono}$ associated to this distortion as $p_{mono}=d Q$, where $d=0.5pm$ is the ferroelectric displacement and $Q$ is the partial charge of the ions. 
We can estimate $Q$ from our DFT calculations as the charge inside the muffin tin sphere of the Ni atoms, this corresponds to $Q=0.9e$ with $e$ the charge of the electron. We then obtain that $p_{mono}=2\cdot10^{-2}$ D, this correspond to an electric polarization of the monolayer of $P_{mono}=5\cdot10^{-13}$ C/m and $\xi\sim10^{-21}$ C from eq. (\ref{polarization}). As a reference, it is interesting to compare this value to the experimental one reported for bulk\cite{PhysRevB.87.014429}. An experimental value for the volumetric electric polarization in bulk of $50\cdot10^{-6}$ C/m$^2$ is reported there. This leads to an electric polarization per layer of $P_{bulk}=4\cdot10^{-14}$ C/m and an electric dipole per layer of $p_{bulk}=10^{-3}$ D.
Interestingly, the estimate we have obtained for the monolayer system overcomes the value reported for the bulk. Most importantly, this highlights that atomic displacements on the order $d=0.5$ pm are not only sizable, but also would provide a ferroelectric order stronger than the bulk value. A more accurate determination of the electric polarization of the monolayer could be performed taking the full microscopic spatial dependence of the charge density, which would account for the partially covalent nature of NiI$_2$. Effectively, this would lead to a renormalization of the effective displaced charge $q$.
Our estimate of the electric polarization of the monolayer is on the typical order of magnitude for spin-driven multiferroic mechanisms\cite{reviewMF2016}.

To understand the contribution of the spin-orbit coupling of each element, we have
further factored the full Hamiltonian in the form

\begin{equation}
\mathcal{H} = 
\mathcal{H}_0 +
\lsoc^{\text{Ni}} \mathcal{H}^{\text{Ni}}_{\text{SOC}} + 
\lsoc^{\text{I}} \mathcal{H}^{\text{I}}_{\text{SOC}}
\end{equation}

where $\lsoc^{\text{Ni}}$, $\lsoc^{\text{I}}$ controls the strength of the SOC in Ni and I, and $\mathcal{H}^{\text{Ni}}_{\text{SOC}}$, $\mathcal{H}^{\text{I}}_{\text{SOC}}$ are the two contributions to the spin-orbit coupling from Ni and I.  
To further demonstrate the dominating role of iodine, we switched off alternatively the SOC contribution of Ni ($\lsoc^{\text{Ni}}=0$ in Fig. \ref{Fig:forces}c) and I ($\lsoc^{\text{I}}=0$ in Fig. \ref{Fig:forces}d) and computed the ferroelectric force as a function of the other one $\lsoc^{\text{I, Ni}}$, respectively. 
In particular, as shown in Fig. \ref{Fig:forces}c, we observe that the ferroelectric force yields nearly identical values to the calculations with full SOC, demonstrating that the dominant contribution to the ferroelectric force arises from the iodine spin-orbit coupling.
In stark contrast, if only the spin-orbit contribution of Ni is included, the ferroelectric force would become much smaller as shown in Fig. \ref{Fig:forces}d.
Indeed, comparing Figs. \ref{Fig:forces}b and \ref{Fig:forces}c, it can be observed that the SOC contribution from the Ni atoms acts in the opposite way to the I contribution for the emergence of a ferroelectric polarization. The previous phenomenology stems from the fact that
SOC of I atoms is one order of magnitude higher than that of Ni atoms, leading to iodine spin-orbit coupling as the dominating contribution to the 2D multiferroic order in NiI$_2$.

Finally, we comment on the implications of the previous phenomenology. First, the ferroelectric polarization is uniquely determined by the spiral $\mathbf{q}$-vector. This implies that in a magnetic domain wall between different $\mathbf{q}$-vectors will lead to an associated domain wall between in the ferroelectric polarizations\cite{domain2009}. 
Second, the magnetic origin of the ferroelectric polarization accounts for the development of ferroelectric polarization right at the magnetic ordering temperatures as observed experimentally\cite{song2021experimental}. 
Therefore, the melting of multiferroicity will be associated to the thermal quench of magnetic order. For a 2D magnetic monolayer with magnetic anisotropy, the melting of magnetic ordering is associated to the thermal excitation of magnetic fluctuations\cite{CrI32017,Burch2018,halides2020,reviewmag2021}. These lead to a critical point at an energy scale controlled by the exchange coupling and the magnetic anisotropy\cite{CrI32017,Torelli2018}. As the parent magnetic ordering is quenched with increasing temperature, so is the ferroelectric ordering, leading to leading to a simultaneous melting of the two orders associated to the multiferroic behavior\cite{reviewMF2009,reviewMF2016}.
Moreover, the energy to flip the ferroelectric polarization will be given by the magnetic exchange, since switching the ferroelectric domain requires switching the magnetic one\cite{reviewMF2016}. However, we have to note that, besides these energetic considerations, the required voltage to create the transition will also depend on the magnetoelectric path followed by the system and the electronic screening of the device.
Importantly, the crucial role of spin-orbit coupling observed in Fig. \ref{Fig:forces} suggests that ferroelastic couplings can be tunable by means of spin-orbit engineering\cite{Fazel2018,Tartaglia2020}.  The strong locking of ferroelectric and magnetic order highlights that conventional strategies of tuning magnetic ordering such as gating\cite{control2018}, strain\cite{PhysRevB.98.144411,Absence2019,strain2020} and exchange proximity\cite{Enhanced2017} can have a dramatic impact on the ferroelectric polarization of the monolayer.  As a consequence, the strong magneto-elastic coupling suggests that twisted multilayers of NiI$_2$ can potentially show emergent moire spiral and ferroelectric orders due to the stacking-induced modulations of the interlayer exchange coupling\cite{stacking2018,stacking2019}.

\section{Conclusions}

We have shown that the multiferroic behavior in monolayer NiI$_2$ stems from the interplay of spin helical order and spin-orbit coupling. In particular, using first-principles calculations, the strong spin-orbit coupling of iodine was shown to dominate the electronic reconstruction leading to the in-plane ferroelectric order.  Interestingly, the strongest charge reconstruction leading to the ferroelectric order emerges around the iodine atoms, which is accompanied by a sizable local magnetization in iodine.  This highlights the fine interplay between spin-orbit coupling and spin non-collinearity as driving forces for multiferroic order, leading to a locking between magnetic helical and ferroelectric orders.  Furthermore, associated with the electronic ferroelectric order, a structural distortion is triggered in the lattice leading to strong magneto-elastic coupling in monolayer NiI$_2$.  Our results show the relativistic origin of multiferroic order in monolayer NiI$_2$, providing a microscopic mechanism accounting for the simultaneous ferroelectric and magnetic ordering.  These results provide a starting point towards controlling multiferroicity in monolayer NiI$_2$ by means of gating, chemical engineering, twist engineering, and proximity effects, ultimately leading to a new family of designer van der Waals multiferroics.

\section*{Acknowledgements}

We acknowledge the computational resources provided by
the Aalto Science-IT project,
and the financial support from the
Academy of Finland Projects No. 331342 and No. 336243,
and the Jane and Aatos Erkko Foundation.
We thank P. Liljeroth, S. Kezilebieke
and V. Pardo for useful discussions.

\bibliography{nii2}

\end{document}